\documentclass[a4paper]{jpconf}
\usepackage{amsmath}
\usepackage{graphicx}
\usepackage{bm}
\begin{document}
\title{Energetic selection of ordered states in a model of the Er$_{2}$Ti$_{2}$O$_{7}$ frustrated pyrochlore XY antiferromagnet}

\author{P A McClarty$^{1}$, S H Curnoe$^{2}$ and M J P Gingras$^{1,3}$}

\address{$^{1}$ Department of Physics and Astronomy, University of Waterloo, Waterloo, Ontario, N2L 3G1, Canada. }
\address{$^{2}$ Department of Physics and Physical Oceanography, Memorial University of Newfoundland, St. John's, NL, A1B 3X7, Canada. }
\address{$^{3}$ Canadian Institute for Advanced Research, Toronto, Ontario, M5G 1Z8, Canada. }

\ead{gingras@gandalf.uwaterloo.ca, curnoe@mun.ca}

\providecommand{\eto}{Er$_{2}$Ti$_{2}$O$_{7}$}
\providecommand{\etos}{Er$_{2}$Ti$_{2}$O$_{7}$ }

\begin{abstract}
We consider the possibility that the discrete long-range ordered states of \etos are selected energetically at the mean field
level as an alternative scenario that suggests selection via thermal fluctuations. We show that nearest neighbour exchange
interactions alone are not sufficient for this purpose, but that anisotropies arising from excited single ion crystal field states in
\eto, together with appropriate anisotropic exchange interactions, can produce the required long range order. However, the effect
of the single ion anisotropies is rather weak so we expect thermal or quantum fluctuations, in some guise, to be ultimately
important in this material. We reproduce recent experimental results for the variation of magnetic Bragg peak intensities as a
function of magnetic field.
\end{abstract}

\section{Introduction}
\label{sec:Introduction}

Frustration is a common feature of magnetic interactions on a pyrochlore lattice. It often has the consequence of introducing a large
degree of degeneracy into the ground state. This might be relieved by weaker interactions with the result that long-range order
sets in at temperatures much smaller than the overall energy scale set by the interactions which, itself, is typically correlated with the
Curie-Weiss temperature. Fluctuations alone can relieve the energetic degeneracy (e.g. \cite{OrderbyDisorder, ChampionHoldsworth})
by a mechanism called order-by-disorder or even select states that do not lie within
the ground state manifold \cite{OrderbyDisorder2, OrderbyDisorder3}. The suppression of the transition temperature compared to the
characteristic energy scale of the interactions can result in quantum fluctuations being more important than thermal fluctuations
in dictating the ordering. Another possibility is that the degeneracy is not relieved in spite of weaker interactions or
fluctuations so that spin liquids or states with unconventional long-range order emerge (see, for example \cite{SpinLiquid,
  QuantumSpinLiquid, Henley}).

The diversity of theoretical possibilities that arises as a consequence of magnetic frustration is paralleled by experiments which have
brought many unusual materials to light. Even among the heavy rare earth titanate pyrochlore
magnets, one can find materials that have no conventional long-range order at the lowest temperatures explored - the spin
ices Ho$_{2}$Ti$_{2}$O$_{7}$ and Dy$_{2}$Ti$_{2}$O$_{7}$ \cite{SpinIceReview}, and the cooperative paramagnets
Tb$_{2}$Ti$_{2}$O$_{7}$ \cite{TTO} and
Yb$_{2}$Ti$_{2}$O$_{7}$ \cite{YTO}, and materials which do exhibit conventional long-range order, but which are puzzling in their
own right: Gd$_{2}$Ti$_{2}$O$_{7}$ \cite{GTO} and Er$_{2}$Ti$_{2}$O$_{7}$ \cite{Champion, Poole,Ruff}. 

\begin{figure}
\includegraphics[width=5cm,clip]{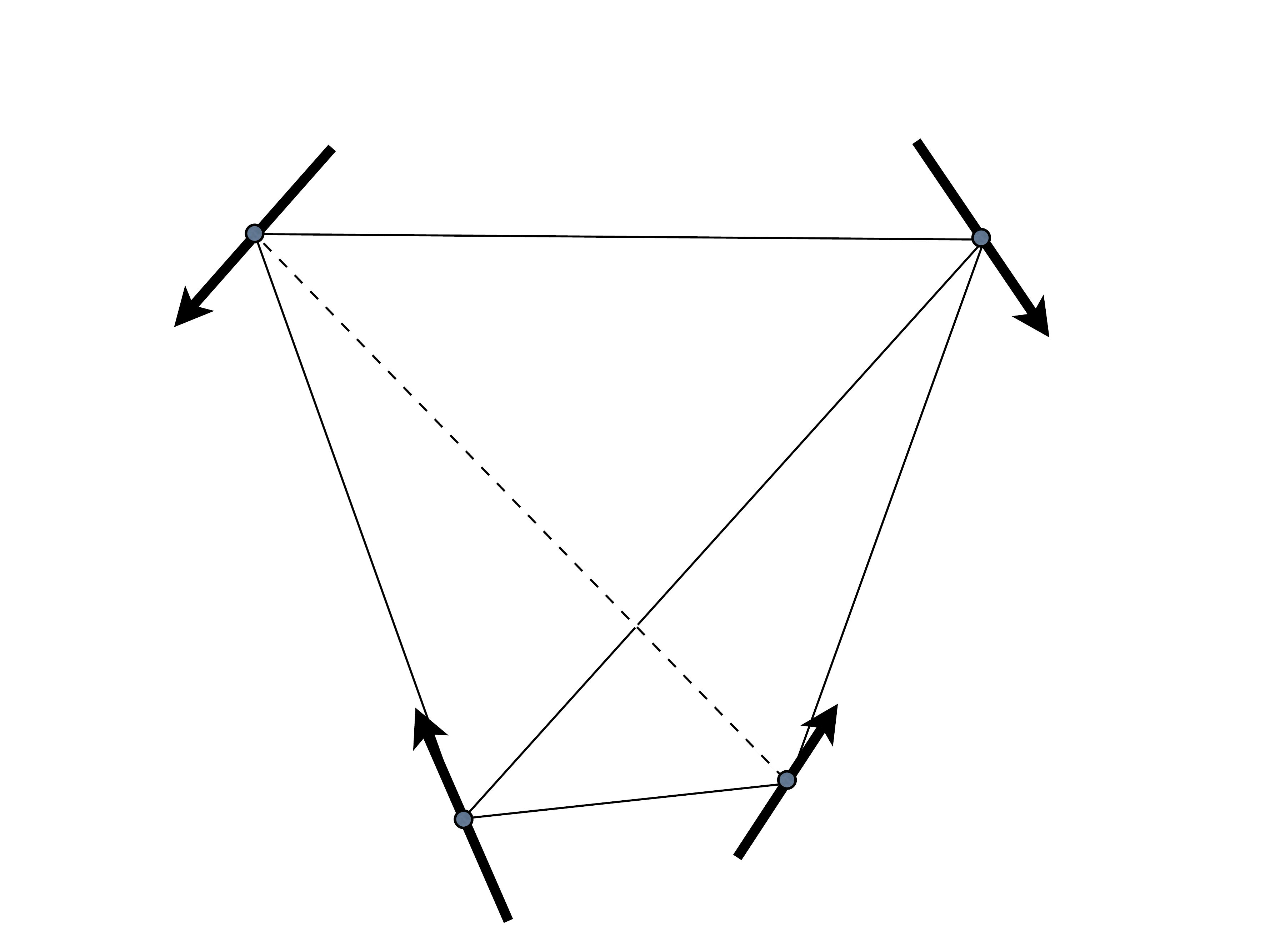}
\hspace{2pc}
\begin{minipage}[b]{3in}
\caption{\label{fig:ETOStates} One ordered domain ($\psi_{2}$ state) of Er$_{2}$Ti$_{2}$O$_{7}$. The spins are in the easy planes
  perpendicular to the local $\langle 1 1 1 \rangle$ directions. The other five possible ground state configurations are obtained
  by discrete rotations and time reversal.}
\end{minipage}
\end{figure}

In this article, we consider the latter material. \eto, which has a Curie-Weiss temperature $\theta_{{\rm CW}}\approx
-15.9$ K \cite{Bramwell}, undergoes a zero-field transition at about $1.2$ K into a phase with ordering
wavevector $\mathbf{q}=0$ \cite{Champion,Poole}. The magnetic structure has been characterized through an analysis of spherical neutron
polarimetry data \cite{Poole}. There are six domains which can be represented by showing the spin configuration on a single
tetrahedron. Figure \ref{fig:ETOStates} represents the spin orientation on a tetrahedral primitive unit cell for a
$\mathbf{q}=0$ ordered state; two others are obtained by discrete global rotations and the
remainder are time reverses of the first three. Henceforth, we refer to these as $\psi_{2}$ states to conform to the notation used
in \cite{Champion} and \cite{Poole}; these states transform according to the magnetic two dimensional irreducible representation
$E_{g}$ based on the octahedral point group $O_{h}$ of the pyrochlore lattice for which the spins are oriented perpendicular to the
local $\langle 1 1 1 \rangle$ directions. The crystal field is responsible for the local $xy$ anisotropy, but the
occurrence of discrete ordered states is not presently understood \cite{ChampionHoldsworth, Champion}. 

An inelastic neutron scattering scan for scattering along the $\mathbf{q}=[22L]$ direction has revealed a gapless or almost gapless
mode in zero field at $\mathbf{q}=[220]$ \cite{Ruff}. As a magnetic field $\mathbf{H}$ along $[ 110 ]$ increases from zero, one finds that
the spectrum is gapped for fields around $|\mathbf{H}|\sim0.5$T and for strong fields $|\mathbf{H}|\sim 5.0$T but that there is a
regime in between where it is apparently gapless. It has been suggested, on this basis that there is a field-driven quantum phase
transition in this material \cite{Ruff}.

In connection with \eto, an easy plane model, with classical spins constrained to lie in the local $\langle 111 \rangle$ planes and interacting via isotropic exchange \cite{ChampionHoldsworth,
  Champion}, has been studied. Energetically, this model has an extensive ground state degeneracy but Monte Carlo
  simulations show that there is a thermally-driven order-by-disorder transition to the $\psi_{2}$ states. However, it is known
  that this cannot be the complete solution to the problem of $\psi_{2}$ selection in \etos \cite{ChampionHoldsworth,
  Champion}. First of all, experimentally, the zero field transition is known to be continuous, whereas the simulations reveal a
  strong first order transition. Another problem is that the Er$^{3+}$ ions have a dipole moment that breaks the degeneracy
  associated with the isotropic exchange, leading to a different $\mathbf{q}=0$ ordering to that seen in \etos
  \cite{PalmerChalker}. Finally, the heat capacity in the low temperature phase varies as $T^{3}$ whereas the model predicts a
  linear variation with temperature.

This paper investigates the conditions under which the selection of $\psi_{2}$ states as ground states may be determined energetically
and at the mean field level in the real compound. We reason as follows: regardless of the
mechanism behind the observed ordering, a six-fold anisotropy must be present in the real system to select the discrete set of
states. Whether this anisotropy is generated ``energetically'', ``entropically'' or ``dynamically'' via quantum fluctuations is
most likely not a meaningful question from an experimental point of view because, in all likelihood, the experimental observables
determined for a material exhibiting true order-by-disorder could be mimicked by an appropriate choice of interactions that
uniquely select the ordered states. Given that the only known order-by-disorder mechanism is problematic and that energetic
selection can be studied, for example, within mean field theory, we adopt the latter approach. The question is then whether such
interactions could be natural in \etos in the sense that they could be justified within a microscopic model. We investigate this
issue in the following.

\section{Selection by exchange interactions}
\label{section:symmetry}

In this section, we look at the problem of whether the most general nearest neighbour interactions, expressed in terms of the
Er$^{3+}$ angular momenta on a pyrochlore, can break the local XY spin rotation symmetry down to the observed ordered states of \etos without any single ion anisotropies. We
show that, in fact, these interactions are not sufficient to select the required states. We start from the nearest neighbour
exchange interactions allowed by symmetry on a pyrochlore lattice which can be found in \cite{Curnoe} and which we present here, in
Table \ref{table:exchange}, with respect to four sets of local axes - one for each sublattice.  The sum of the terms in each of the four columns of the Table is a linear
combination of bilinear operators that is invariant under the point group of the pyrochlore lattice $O_{h}$. We denote the four
invariants by $X_{p}$ for $p=1,\ldots,4$. The isotropic exchange interaction, nearest neighbour dipole-dipole interaction,    
$\mathbf{J}_{i}\cdot\mathbf{J}_{j} - 3(\mathbf{J}_{i}\cdot\mathbf{\hat{r}}_{ij})(\mathbf{J}_{j}\cdot\mathbf{\hat{r}}_{ij})$, where $\mathbf{\hat{r}}_{ij}$ are unit vectors connecting nearest neighbours, and the Dzyaloshinskii-Moriya (DM) interaction,
$\mathbf{D}\cdot(\mathbf{J}_{i}\times\mathbf{J}_{j})$ \cite{Elhajal,Elhajal2}, can be written as linear combinations
of these invariants. For example, isotropic exchange $\sum_{\langle i,j \rangle}\mathbf{J}_{i}\cdot\mathbf{J}_{j}=
\sum_{p=1}^{4}X_{p}$. Ordinarily the dipole-dipole interaction is long-ranged; because we cut it off beyond nearest neighbours, which
is not unreasonable for modelling a zero moment $\mathbf{q}=0$ ordering, we refer to it as the pseudo-dipole interaction.

We can express these exchange interactions in terms of bilinears that are the products of basis vectors of the point group. For
\eto, the irreducible representation that goes critical at the transition temperature is the two dimensional irreducible
representation $E_{g}$ \cite{Champion} which has orthogonal basis vectors that we denote by $J_{E_{+}}$ and $J_{E_{-}}$ from which the order parameter can be
constructed. These basis vectors are written explicitly in terms of $J^{\alpha}$ components in equation \ref{eqn:EPEM} in the Appendix. One can show that these
basis vectors appear in the symmetry allowed exchange interactions $X_{p}$ only in the combination $J_{E_{+}}J_{E_{-}}$ (see
Appendix). The crucial point of the argument is that this exchange term does {\it not} distinguish any directions in the easy
planes of the four sublattices so, in the thermodynamic limit, one would expect a continuous symmetry breaking provided the coefficient of
$J_{E_{+}}J_{E_{-}}$ in the Hamiltonian is negative. This situation is captured by a Landau theory based on the $E_{g}$ irreducible
representation to quartic order in the order parameter. The order parameter for \etos is the expectation value of
$\alpha_{E_{+}}J_{E_{+}}+\alpha_{E_{-}}J_{E_{-}}$ for $\alpha_{E_{+}}=\alpha_{E_{-}}=1$ and then five other values related to the
first by symmetry. These are minima of a Landau theory to sixth order in the invariants of $E_{g}$ (see, for example,
\cite{Sergienko}). 

The conclusion of this section is that nearest neighbour exchange interactions allowed by symmetry on a
pyrochlore cannot, in the absence of single ion anisotropies, break the symmetry uniquely down to the $\psi_{2}$ states. In the
next section, we explore a mechanism for the generation of sixth order terms in the Landau theory for \eto. 
\begin{table}
\caption{\label{table:exchange} Exchange interactions over a single tetrahedron expressed in terms of local coordinates for
  each magnetic ion. The coordinate system is described in the Appendix. Invariant $X_{p}$ (for each $p$) is the sum over terms in its
  column. Term $\mathbf{J}_{i} \cdot \mathbf{J}_{j}$ for given $i$ and $j$ is the sum over terms in its row. $\epsilon = \exp(2\pi
  i/3)$. $\mbox{h.c.}$ stands for Hermitian conjugate.} 
\begin{center}
\begin{tabular}{| c | c | c | c | c |}
\br
Term & $X_{1}$ & $X_{2}$ & $X_{3}$ & $X_{4}$ \\
\mr
$\mathbf{J}_{1}\cdot \mathbf{J}_{2}$ & $-\frac{1}{3}J^{z}_{1}J^{z}_{2}$ & $-\frac{\sqrt{2}}{3}(J^{z}_{1}J^{+}_{2} +
J^{+}_{1}J^{z}_{2}) + \mbox{h.c.}$ & $\frac{1}{3}J^{+}_{1}J^{+}_{2} +
\mbox{h.c.}$ & $-\frac{1}{6}J^{+}_{1}J^{-}_{2} + \mbox{h.c.}$ \\
$\mathbf{J}_{1}\cdot \mathbf{J}_{3}$ & $-\frac{1}{3}J^{z}_{1}J^{z}_{3}$  &
 $-\frac{\sqrt{2}}{3}\epsilon(J^{z}_{1}J^{+}_{3} + J^{+}_{1}J^{z}_{3}) + \mbox{h.c.}$ & $\frac{1}{3}\epsilon^{*}J^{+}_{1}J^{+}_{3} + \mbox{h.c.}$ & $-\frac{1}{6}J^{+}_{1}J^{-}_{3} + \mbox{h.c.}$ \\
$\mathbf{J}_{1}\cdot \mathbf{J}_{4}$ & $-\frac{1}{3}J^{z}_{1}J^{z}_{4}$  & $-\frac{\sqrt{2}}{3}\epsilon^{*}(J^{z}_{1}J^{+}_{4} +
J^{+}_{1}J^{z}_{4}) + \mbox{h.c.}$ & $\frac{1}{3}\epsilon J^{+}_{1}J^{+}_{4} + \mbox{h.c.}$ & $-\frac{1}{6}J^{+}_{1}J^{-}_{4} + \mbox{h.c.}$ \\
$\mathbf{J}_{2}\cdot \mathbf{J}_{3}$ & $-\frac{1}{3}J^{z}_{2}J^{z}_{3}$  & $-\frac{\sqrt{2}}{3}\epsilon^{*}(J^{z}_{2}J^{+}_{3} +
J^{+}_{2}J^{z}_{3}) + \mbox{h.c.}$ & $\frac{1}{3}\epsilon J^{+}_{2}J^{+}_{3} + \mbox{h.c.}$ & $-\frac{1}{6}J^{+}_{2}J^{-}_{3} + \mbox{h.c.}$ \\
$\mathbf{J}_{2}\cdot \mathbf{J}_{4}$ & $-\frac{1}{3}J^{z}_{2}J^{z}_{4}$  & $-\frac{\sqrt{2}}{3}\epsilon(J^{z}_{2}J^{+}_{4} +
J^{+}_{2}J^{z}_{4}) + \mbox{h.c.}$ & $\frac{1}{3}\epsilon^{*} J^{+}_{2}J^{+}_{4} + \mbox{h.c.}$ & $-\frac{1}{6}J^{+}_{2}J^{-}_{4} + \mbox{h.c.}$ \\
$\mathbf{J}_{3}\cdot \mathbf{J}_{4}$ & $-\frac{1}{3}J^{z}_{3}J^{z}_{4}$  & $-\frac{\sqrt{2}}{3}(J^{z}_{3}J^{+}_{4} +
J^{+}_{3}J^{z}_{4}) + \mbox{h.c.}$ & $\frac{1}{3}J^{+}_{3}J^{+}_{4} + \mbox{h.c.}$ &
 $-\frac{1}{6}J^{+}_{3}J^{-}_{4} + \mbox{h.c.}$ \\
\br
\end{tabular}
\end{center}
\end{table}

\section{Inclusion of in-plane anisotropies}
\label{section:MFT}

We mentioned in Section \ref{sec:Introduction} that, with isotropic exchange interactions, the pyrochlore with XY anisotropy has an extensive ground state
degeneracy \cite{ChampionHoldsworth}, and that the discrete $\psi_{2}$ states belong to this continuous set of ground states. Then, the
inclusion of a six-fold anisotropy, with minima along the spin directions of the $\psi_{2}$ states, must break the degeneracy
exclusively down to the six $\psi_{2}$ domains observed in \eto. Single ion anisotropies were omitted in the symmetry arguments of
Section \ref{section:symmetry}. With classical spins, the relevant single ion anisotropy is $J_{+}^{6}+J_{-}^{6} \sim \cos(6\phi)$, where
$\phi$ is an angle in the local $\langle 111\rangle$ plane.  This is proportional to the Racah operator $O_{6}^{6}$ which appears in the crystal field
Hamiltonian \cite{Rosenkranz}. So, although the lowest crystal field doublet has almost perfect XY degeneracy, one might expect that the mixing of excited
crystal field levels into the ground state induced by the interactions between the angular momenta could perhaps introduce a six-fold
anisotropy of the correct modulation to break the symmetry energetically.  This would fulfil the requirement that the Hamiltonian
be well-motivated from a microscopic point of view. We note, however, that recent inelastic neutron scattering experiments on
\etos \cite{Ruff} do not resolve a gap to spin excitations which should be present if the symmetry is broken in the ground
state. However, as we point out below, one would expect this gap, if it exists, to be small.

To investigate whether the excited crystal field levels can introduce a six-fold anisotropy with the correct modulation, we have
studied a microscopic Hamiltonian $H=H_{{\rm cf}}+H_{{\rm ex}}$ where
\begin{align}
H_{\mbox{cf}} & = \sum_{i} B^{2}_{0}O^{2}_{0,i} + B^{4}_{0}O^{4}_{0,i} +
  B^{4}_{3}O^{4}_{3,i} +  B^{6}_{0}O^{6}_{0,i} +  B^{6}_{3}O^{6}_{3,i} + B^{6}_{6}O^{6}_{6,i} \\
H_{\mbox{ex}} & = \frac{1}{2}\sum_{\mu,\nu}\sum_{a,b}\sum_{\alpha,\beta}\mathcal{J}_{ab}^{\alpha\beta}(\mathbf{R}_{\mu\nu}) J_{a}^{\alpha}(\mathbf{R}_{\mu})J_{b}^{\beta}(\mathbf{R}_{\nu}). 
\end{align}
The Greek indices are angular momentum components, $a$ and $b$ are sublattice indices (ranging from $1$ to $4$) and $\mathbf{R}_{\mu}$ is a
primitive lattice vector with
$\mathbf{R}_{\mu\nu}\equiv\mathbf{R}_{\mu}-\mathbf{R}_{\nu}$; index $i$ runs
over all magnetic ions.
The crystal field parameters ${ B^{l}_{m} }$ are rescaled from those of Ho$_{2}$Ti$_{2}$O$_{7}$ \cite{Rosenkranz}, and the choice
of nearest neighbour exchange couplings ${\omega_{p}}$ is guided by the requirement that, in the phase diagrams to be presented
below, the order of magnitude of the couplings should be consistent with the experimental
Curie-Weiss temperature \cite{Bramwell} which we have fitted by computing the uniform susceptibility in the random phase
approximation (RPA) \cite{RareEarths}, and in the paramagnetic phase, from the Hamiltonian given above. We rewrite $H_{{\rm ex}}$ on a
single tetrahedron as $\sum_{p=1}^{4}\omega_{p}X_{p}$, hence constraining the elements of $\mathcal{J}_{ab}^{\alpha\beta}$.

To proceed, we carry out a mean-field decoupling of $J_{a}^{\alpha}(\mu)J_{b}^{\beta}(\nu)$ by writing it as
\begin{equation} J_{a}^{\alpha}(\mu)\langle J_{b}^{\beta}(\nu)\rangle
  + J_{b}^{\beta}(\nu)\langle J_{a}^{\alpha}(\mu)\rangle - \langle J_{a}^{\alpha}(\mu)\rangle\langle J_{b}^{\beta}(\nu)\rangle 
 + (J_{a}^{\alpha}(\mu)-\langle J_{a}^{\alpha}(\mu)\rangle) (J_{b}^{\beta}(\nu)-\langle J_{b}^{\beta}(\nu)\rangle  )
\end{equation} 
and omitting the last term which quantifies the fluctuations. The angled brackets indicate a thermal average $\langle A \rangle
\equiv \frac{1}{Z} \mbox{Tr} \left\{Ae^{-\beta H} \right\}$, and we have abbreviated $J_{a}^{\alpha}(\mathbf{R}_{\mu})$ by
$J_{a}^{\alpha}(\mu)$.

The original interacting problem is reduced to a single ion problem
\begin{equation} H_{{\rm MF}} = H_{{\rm cf}} + \sum_{\mu,a,\alpha} \left(J_{a}^{\alpha}(\mu) - \frac{1}{2}\langle J_{a}^{\alpha}(\mu)\rangle
\right)h_{a}^{\alpha}(\mu) \label{eqn:HMF} \end{equation}
where $h_{a}^{\alpha}$ is an effective magnetic field of the form $ h_{a}^{\alpha}(\mu) \equiv \sum_{\nu,b,\beta}
\mathcal{J}_{ab}^{\alpha\beta}(\mathbf{R}_{\mu\nu})\langle J_{b}^{\beta}(\nu)\rangle$. 

Mean field solutions for a cubic unit cell and examination of the RPA susceptibility show that, for the parameters we have explored,
the ordering wavevector is $\mathbf{q}=0$. So, for a qualitative investigation, we can, at this time, restrict our attention to a
single tetrahedron.

\begin{figure}[h]
\begin{minipage}{18pc}
\includegraphics[width=18pc]{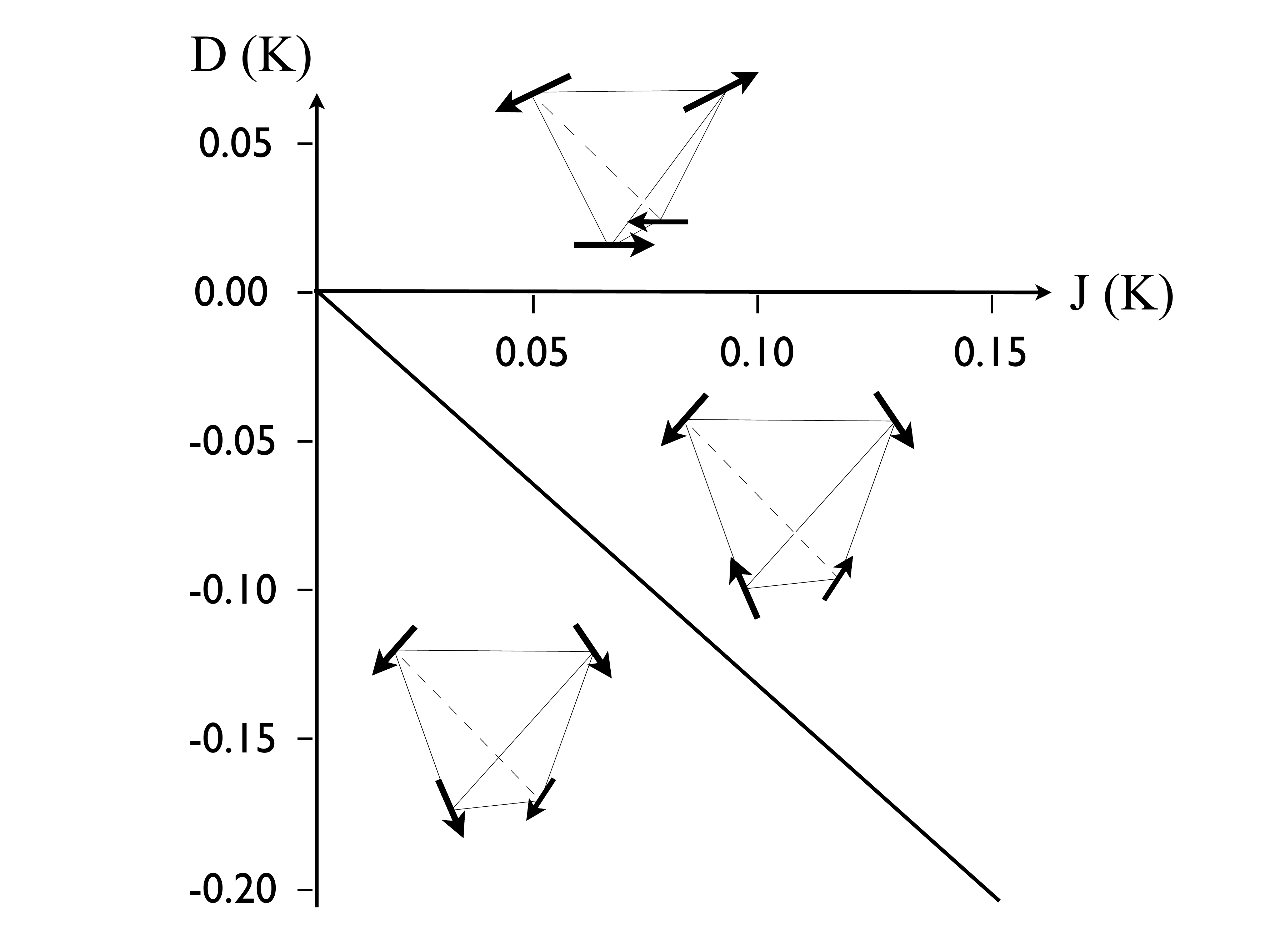}
\caption{\label{fig:phase1} Phase diagram for isotropic exchange $J$ and pseudo-dipole $D$ couplings. Three different
  $\mathbf{q}=0$ phases arise as indicated. From bottom to top, there a configuration with a net moment, then the $\psi_{2}$
  state, and then the state reported in \cite{PalmerChalker}.}
\end{minipage}\hspace{2pc}
\begin{minipage}{18pc}
\includegraphics[width=18pc]{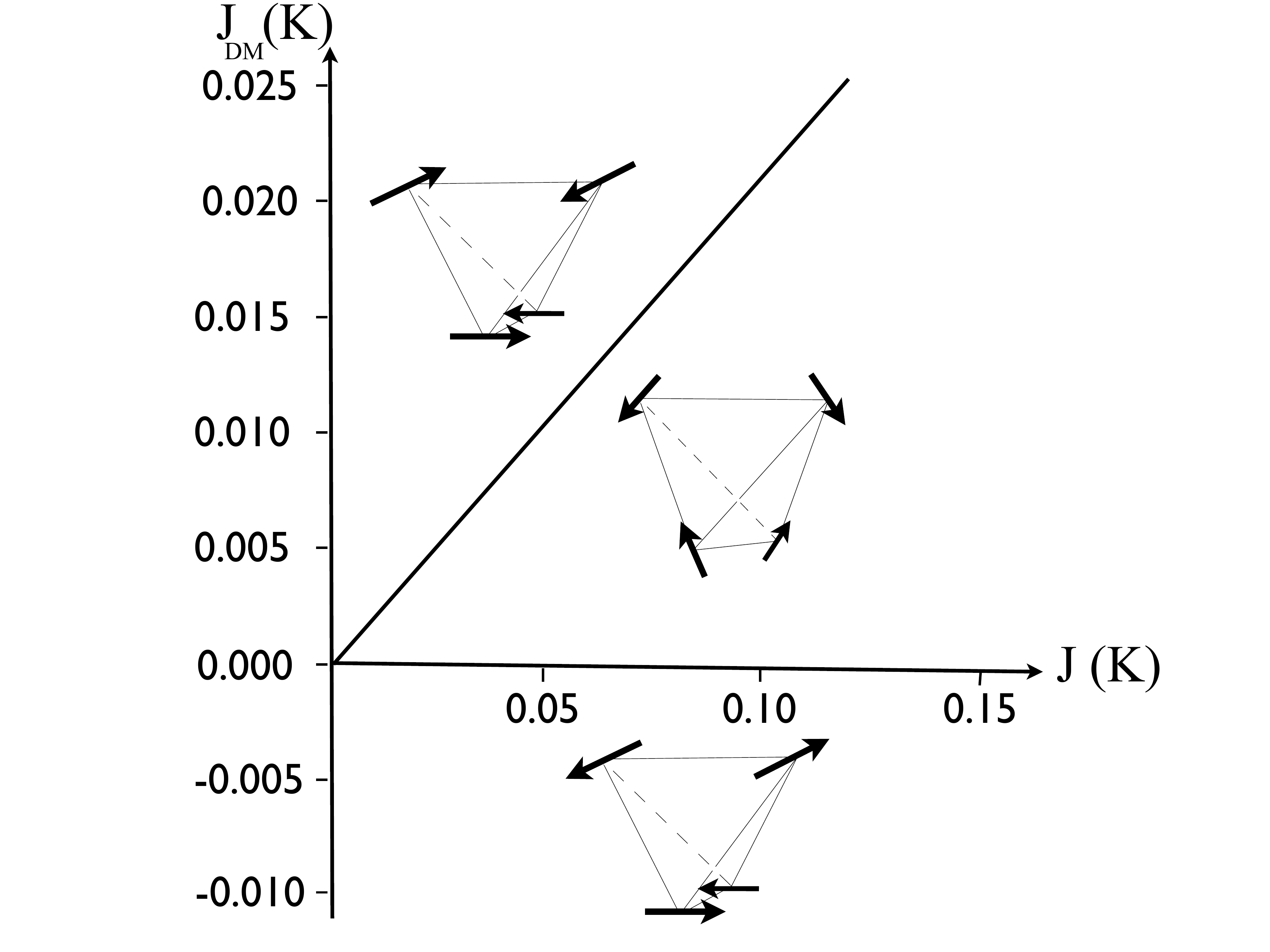}
\caption{\label{fig:phase2} Phase diagram for isotropic exchange $J$ and Dzyaloshinskii-Moriya interactions $J_{\rm DM}$. From bottom to
  top, there is a state familiar for dipole interactions \cite{PalmerChalker}, then the $\psi_{2}$ state and then a $\psi_{3}$
  state \cite{Champion}.}
\end{minipage}
\end{figure}

Starting from random choices for the twelve expectation values $\langle J_{a}^{\alpha}\rangle$, the Hamiltonian is iterated to
self-consistency for different couplings. We have found that the six $\psi_{2}$ domains are solutions to the mean field equations
for some regions of parameter space. Figures \ref{fig:phase1} and \ref{fig:phase2} show phase diagrams including each showing such
a region - isotropic exchange and pseudo-dipole interactions, and isotropic exchange with DM interactions respectively. The self-consistent
solutions are indicated on the phase diagrams. As expected, we find the $\mathbf{q}=0$ states reported in \cite{PalmerChalker}
with isotropic exchange and positive pseudo-dipole coupling. Though the physical dipole-dipole coupling is positive, we find the
\etos ordered states only when the pseudo-dipole coupling is negative in this case. We find the same $\psi_{2}$ states with
positive DM coupling and isotropic exchange for $J<J_{\rm DM}$ (see Fig. \ref{fig:phase2}). We have also found couplings
$\chi_{1}>0$ and $\chi_{2}<0$, given in the Appendix, such that $\psi_{2}$ order occurs for the positive
pseudo-dipole coupling of $D=0.02$ K appropriate to \eto \cite{Champion}.

\begin{figure}[h]
\begin{minipage}{18pc}
\includegraphics[width=18pc]{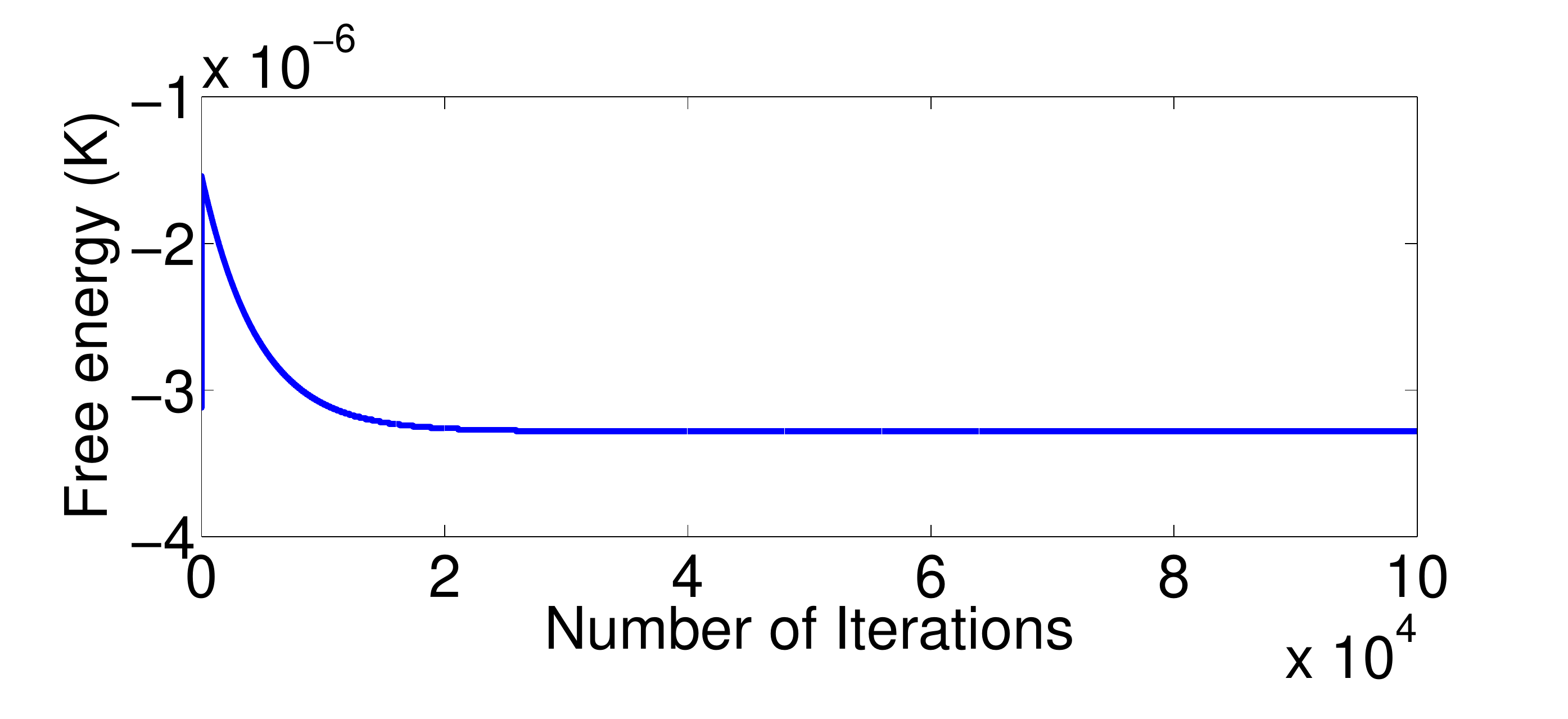}
\caption{\label{fig:IterationFE} Variation of the free energy per tetrahedron upon iteration with $J=0.1$\nolinebreak\ K and
  $D=-0.02$\nolinebreak \
  K. $T=0.3$\nolinebreak\ K.}
\end{minipage}\hspace{2pc}
\begin{minipage}{18pc}
\includegraphics[width=18pc]{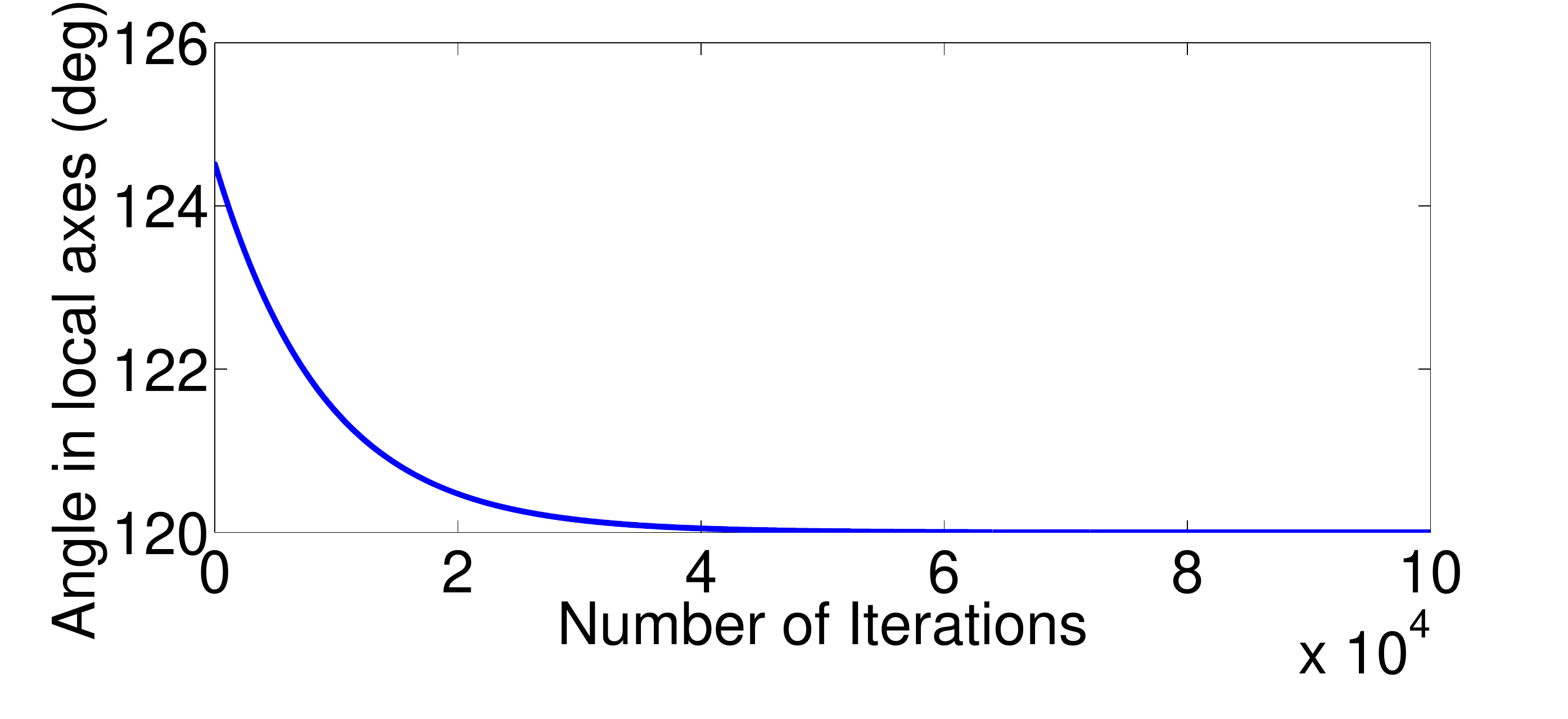}
\caption{\label{fig:IterationAngle} Convergence of the angle upon iteration with mean field parameters $J=0.1$ K and $D=-0.02$
  K. $T=0.3$\nolinebreak\ K.}
\end{minipage}
\end{figure}

The progress of iterations of the mean field equations to self-consistency sheds some light on the robustness of the
energetic selection mechanism. Figure \ref{fig:IterationFE} shows the free energy per site plotted against the number of
iterations when the couplings are such that the $\psi_{2}$ states are selected. The first few iterations have been cut off to see
the scale of the trend. An accompanying plot, Figure \ref{fig:IterationAngle}, shows the variation in the angle of
$\langle \bm{J} \rangle$ in the local $\langle 111 \rangle$ plane. The spins settle into their respective $xy$ planes within a few
iterations but around $O(10^{4})$ iterations are required for the spins to converge to the $\psi_{2}$ angles. The variation in the
free energy across this interval is $O(10^{-6})$ K. This is in marked contrast to the case of isotropic exchange and positive pseudo-dipole
coupling where the convergence is complete within $O(10^{1})$ iterations corresponding to energetic selection without any need for
single ion anisotropies within the local $xy$ planes coming from excited crystal field levels.

This result can be understood in terms of a mean field treatment of the quantum corrections to the Hamiltonian obtained by
projecting onto the space spanned by the ground state crystal field doublet (see, for instance, \cite{Molavian}). Such
corrections are computed perturbatively in powers of $J/\Delta$ where $J$ is the characteristic energy scale of the exchange, which
is about $0.1$ K for \eto, and $\Delta\sim 10^{2}$ K is the scale of the crystal field splitting \cite{Champion}. These terms
involve admixing with excited crystal field states which are responsible for bringing the six-fold anisotropy into the
mean field solutions. The smallness of $J/\Delta$ and the fact that six-fold anisotropy cannot be generated to lowest order in the
perturbative corrections implies that the selection of $\psi_{2}$ states is very weak within the mechanism considered here.

\section{Bragg scattering in a magnetic field}

\begin{figure}[h]
\begin{minipage}{18pc}
\includegraphics[width=18pc]{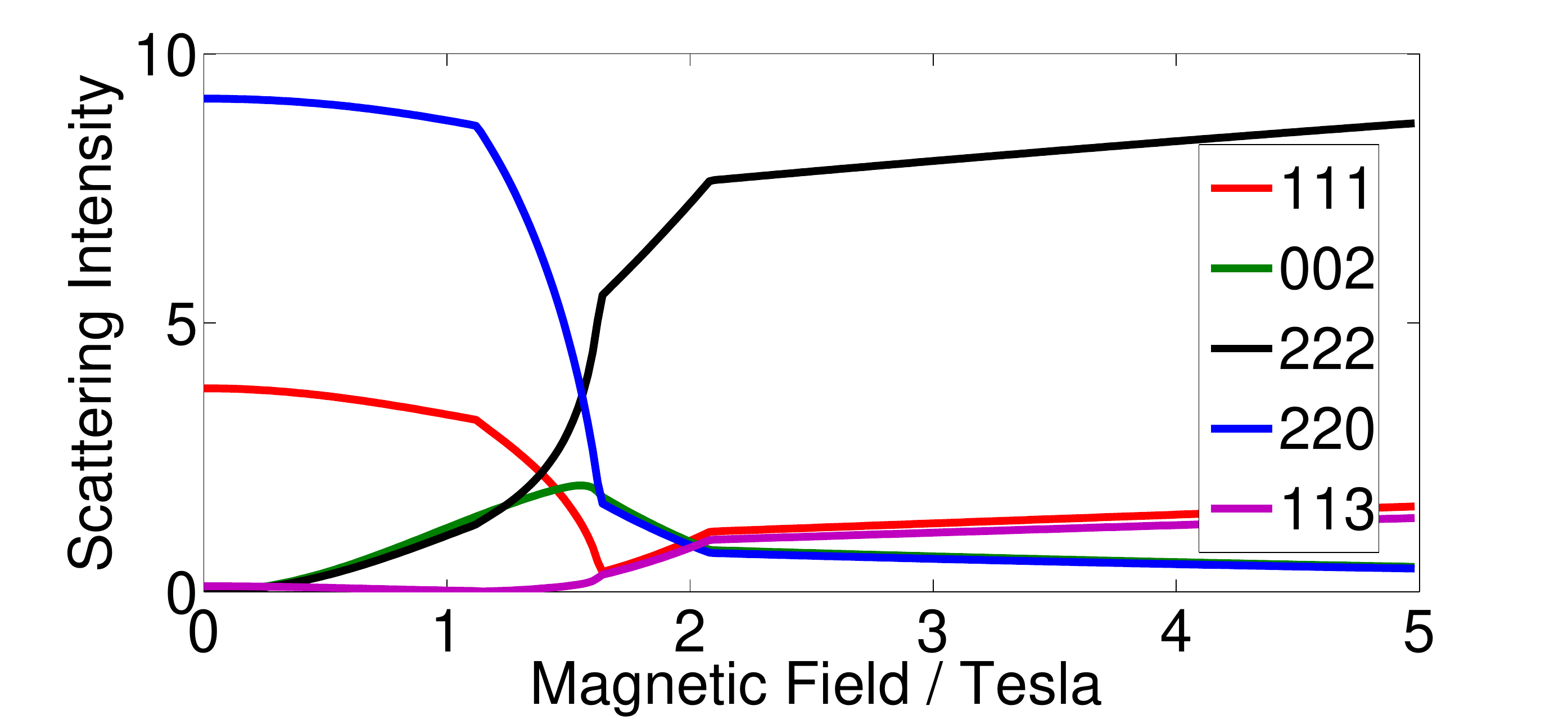}
\caption{\label{fig:MFT_Int} Bragg peak intensities from mean field theory with isotropic exchange $J=0.1$ K and dipole-dipole
  coupling $D=-0.02$ K. $T=0.3$ K.}
\end{minipage}\hspace{2pc}
\begin{minipage}{17pc}
\includegraphics[width=17pc]{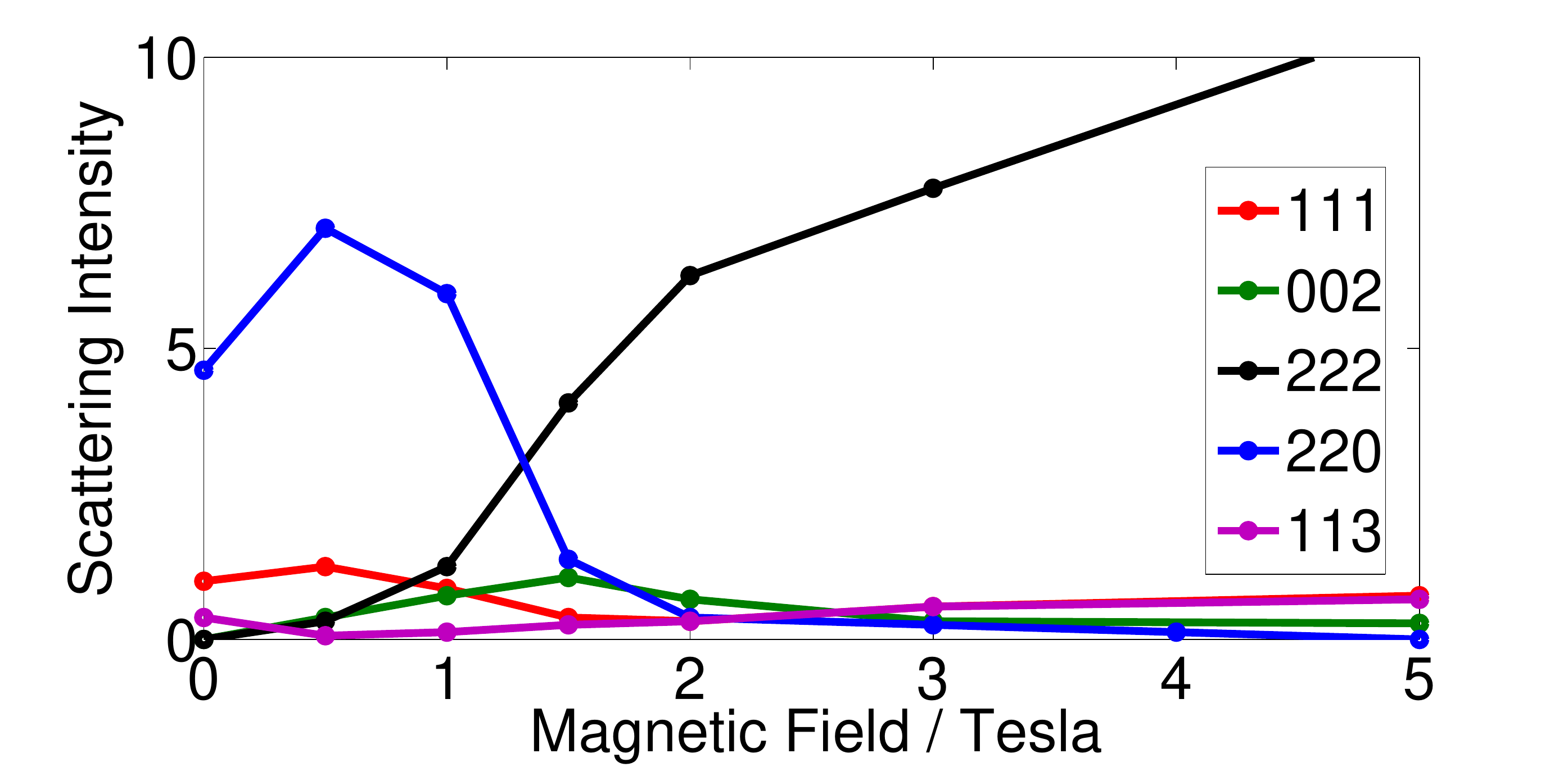}
\caption{\label{fig:Exp_Int} Experimental magnetic Bragg peak intensities for \etos in a $[1,1,0]$ field taken from \cite{Ruff}.
\\}
\end{minipage}
\end{figure}

We have studied the effect on the elastic neutron scattering pattern of a magnetic field applied to the material in its low temperature
ordered state. Following experiments by Ruff and co-workers \cite{Ruff} in which the intensities of five Bragg peaks were measured
as the field strength was varied, we have applied a field in the crystallographic $[110]$ direction. A Zeeman term  $H_{Z} =
-\mu_{B}g\mathbf{J}\cdot\mathbf{H}$ was added to $H_{\rm MF}$ (Equation \ref{eqn:HMF}). The field was increased incrementally starting from one zero
field $\psi_{2}$ domain and the self-consistent solution was found for each field. Figure \ref{fig:MFT_Int} is the mean field
result and Figure \ref{fig:Exp_Int} are the experimental intensities taken from \cite{Ruff}. In the experimental range
$|\mathbf{H}|=0$ T to $|\mathbf{H}|=5$ T, there is good qualitative agreement between the two sets of intensities for the chosen
couplings. The maximum $|\mathbf{H}|=5$ T is sufficiently small that the spins remain in their respective $xy$ planes.  The change
in the spin configuration coincides with the picture presented in \cite{Ruff} that has spins along chains in the field direction
maintaining antiferromagnetic correlations before rotating together into the field direction. Most of the spin rotation happens in
a narrow band around $1.5$ T in both the experiment and in the theory. However, one can change the band of fields for which this
sharp feature is seen by varying the couplings (of course, with the condition that the $\psi_{2}$ states are the $\mathbf{H}=0$ ordered states).

\section{Discussion}

We have asked in this article whether models for \etos can be written down that are microscopically well-motivated and which
energetically select the experimentally observed $\psi_{2}$ states. This approach stands in contrast to earlier investigations
that focus on fluctuation-induced ordering \cite{ChampionHoldsworth,Champion}. We have found that there are indeed such models
with crystal field terms and suitable symmetry-allowed bilinear interactions between the nearest neighbour angular momenta. We
have shown that models with only bilinear interactions cannot select the required states. The same is true for local $\langle 111
\rangle$ plane $xy$ models with these bilinear interactions. The introduction of crystal field terms
does allow for $\psi_{2}$ states to be selected for certain regions in the space of couplings - the interaction-induced admixing of the ground doublet with excited crystal field levels being the source of the necessary six-fold anisotropy. However, because the energy
scale of the crystal field splitting is very much greater than the characteristic scale of the interactions, the selection of
these discrete ground states is extremely weak. Even if the exchange interactions were such that $\psi_{2}$ long-range ordering
would occur, one would expect the transition temperature $T_{c}$ to be very much smaller than the experimental $T_{c}$ provided
that an order-by-disorder mechanism does not select the $\psi_{2}$ states. In view of these results, and the problems with the
isotropic exchange order-by-disorder mechanism that were outlined in the Introduction, we tentatively conclude that neither of
these proposals is sufficient on its own. It is conceivable, instead, that they complement one another - that the microscopic
model might weakly select the $\psi_{2}$ states and that fluctuations have the effect of raising the transition temperature. One
may ask whether order-by-disorder could operate with a continuous transition rather than a strongly first order transition in
these circumstances. It might also be that quantum fluctuations must be considered to find the correct answer to this question as
suggested in \cite{Champion}.

One success of our mean field theory approach is that we have been able to reproduce the experimental variation of the magnetic
Bragg peaks intensities in a magnetic field in the $[110]$ direction. Provided we choose the interactions correctly, in the range
of fields where the Bragg intensities change most sharply, the theory matches the experimental field strength. The theory does not reproduce
the upturn in intensities between $0$ T and $0.5$ T however; the mean field theory selects two domains out of the six in very weak
fields ($|\mathbf{H}|\ll 0.5$ T). Two interpretations for this upturn have been suggested in the literature: the first is that it
is due to the selection of preferred domains \cite{Champion}. The second interpretation comes from the observation that a broad
component to the scattering is suppressed for weak fields and that this coincides with the upturn in the main scattering component
to conserve the intensity. It has been suggested that quantum fluctuations are responsible for the broader scattering peaks and
that these are suppressed rapidly in weak fields \cite{Ruff}. Such an effect may lend weight to an investigation into zero field
ordering in \etos induced by quantum effects. 
 
\subsection{Acknowledgments}

We would like to acknowledge support from the NSERC and the CRC program (Tier 1, MJPG).

\appendix

\section*{Appendix}

\setcounter{section}{1}

The basis of four magnetic ions with respect to face centred cubic system is $\mathbf{r}_{1} = \frac{a}{4}(0,0,0)$,
$\mathbf{r}_{2} = \frac{a}{4}(1,1,0)$, $\mathbf{r}_{3} = \frac{a}{4}(1,0,1)$ and $\mathbf{r}_{4} = \frac{a}{4}(0,1,1)$
where $a$ is the edge length of the cubic unit cell. The local coordinate system is chosen with local $z$ axes in the $\langle
111\rangle$ directions. The $x$ axes are chosen as follows $\mathbf{x}_{1} = \frac{1}{\sqrt{6}}(1,1,-2)$, $\mathbf{x}_{2} =
\frac{1}{\sqrt{6}}(-1,-1,-2)$, $\mathbf{x}_{3} = \frac{1}{\sqrt{6}}(-1,1,2)$ and $\mathbf{x}_{4} = \frac{1}{\sqrt{6}}(1,-1,2)$ and
the axes are right-handed. 

The basis vectors for the different irreducible representations of $O_{h}$ are given below. The four local $z$ components of the
angular momenta are decomposed into $A_{2}\oplus T_{1}$ and the remaining components are $E \oplus T_{1} \oplus T_{2}$.  
\begin{align} 
& J_{E_{+}} = J^{+}_{1} + J^{+}_{2} + J^{+}_{3} + J^{+}_{4} &  
& J_{E_{-}} = J^{-}_{1} + J^{-}_{2} + J^{-}_{3} + J^{-}_{4} \label{eqn:EPEM} \\  
& J_{A_{2}} = J^{z}_{1} + J^{z}_{2} + J^{z}_{3} + J^{z}_{4} \\ 
& J^{x}_{T_{1,1}} = J^{z}_{1} - J^{z}_{2} - J^{z}_{3} + J^{z}_{4} & 
& J^{y}_{T_{1,1}} = J^{z}_{1} - J^{z}_{2} + J^{z}_{3} - J^{z}_{4}  \\
& J^{z}_{T_{1,1}} = J^{z}_{1} + J^{z}_{2} - J^{z}_{3} - J^{z}_{4} \\ 
& J^{x}_{T_{1,2}} = \frac{1}{2}\epsilon^{*} (J^{+}_{1} - J^{+}_{2} - J^{+}_{3} + J^{+}_{4}) + \mbox{h.c.} &
& J^{y}_{T_{1,2}} = \frac{1}{2}\epsilon (J^{+}_{1} - J^{+}_{2} + J^{+}_{3} - J^{+}_{4}) + \mbox{h.c.} \\
& J^{z}_{T_{1,2}} = J^{x}_{1} + J^{x}_{2} - J^{x}_{3} - J^{x}_{4} \\ 
& J^{x}_{T_{2}} = -\frac{i}{2}\epsilon^{*} (J^{+}_{1} - J^{+}_{2} - J^{+}_{3} + J^{+}_{4}) + \mbox{h.c.} &
& J^{y}_{T_{2}} = -\frac{i}{2}\epsilon (J^{+}_{1} - J^{+}_{2} + J^{+}_{3} - J^{+}_{4}) + \mbox{h.c.} \\
& J^{z}_{T_{2}} = J^{y}_{1} + J^{y}_{2} - J^{y}_{3} - J^{y}_{4}
\end{align}
The appearances of the $T_{1}$ irreducible representation have labels to distinguish them: $T_{1,1}$ and $T_{1,2}$. The phase
$\epsilon = \exp(2\pi i/3)$.  The exchange invariants, given in Table \ref{table:exchange}, can be written in terms of these basis
vectors as follows
\begin{align*} 
X_{1} & = -\frac{1}{8}J_{A_{2}}^{2} + \frac{1}{24}\mathbf{J}_{T_{1,1}}^{2} &
X_{2} & = -\frac{\sqrt{2}}{3}\mathbf{J}_{T_{1,1}}\cdot \mathbf{J}_{T_{1,2}}\\
X_{3} & = \frac{1}{6}\mathbf{J}_{T_{1,2}}^{2} - \frac{1}{6}\mathbf{J}_{T_{2}}^{2} &
X_{4} & = -\frac{1}{8}J_{E_{+}}J_{E_{-}} + \frac{1}{24}\mathbf{J}^{2}_{T_{1,2}} + \frac{1}{24}\mathbf{J}^{2}_{T_{2}} 
\end{align*}
so, as stated in Section \ref{section:symmetry}, the only appearance of the basis vectors of the two dimensional
irreducible representation $E_{g}$, which control the symmetry breaking to the $\psi_{2}$ states, is in the combination
$J_{E_{+}}J_{E_{-}}$ which does not lead {\it uniquely} to the $\psi_{2}$ states, as outlined in Section \ref{section:symmetry}. 

In Section \ref{section:MFT}, we make reference to invariants $\{\chi\}$, which are linear combinations of those in Table \ref{table:exchange}, $\chi_{1} \equiv 2X_{1} + \frac{1}{2}X_{2} + \frac{1}{2}X_{3} + 2X_{4}$ and $\chi_{2} \equiv -X_{1} + \frac{1}{2}X_{2}
  + \frac{1}{2}X_{3} - X_{4}$, which have the property that $\chi_{1}+\chi_{2}$ is the isotropic exchange on a tetrahedron.


\section*{References}

\begin{thebibliography}{9}

\bibitem{OrderbyDisorder} Bramwell S T, Gingras M J P, Reimers J N 1994 {\it J. Appl. Phys.} {\bf 75} 5523

\bibitem{ChampionHoldsworth} Champion J D M and Holdsworth P C W 2004 {\it J. Phys.: Condens. Matter} {\bf 16} S665

\bibitem{OrderbyDisorder2} Pinettes C, Canals B and Lacroix C 2002 {\it Phys. Rev. B} {\bf 66} 024422

\bibitem{OrderbyDisorder3} Chern G-W, Moessner R and Tchernyshyov O 2008 {\it Preprint} arXiv:0803.2332 (to appear in Phys. Rev. B)

\bibitem{SpinLiquid} Moessner R and Chalker J T 1998 {\it Phys. Rev. Lett.} {\bf 80} 2929

\bibitem{QuantumSpinLiquid} Hermele M, Fisher M P A and Balents L 2004 {\it Phys. Rev. B} {\bf 69} 064404

\bibitem{Henley} Henley C L 2005 {\it Phys. Rev. B} {\bf 71} 014424

\bibitem{SpinIceReview} Bramwell S T and Gingras M J P 2001 {\it Science} {\bf 294} 1495

\bibitem{TTO} Enjalran M {\it et al.} 2004 {\it J. Phys.: Condens. Matter} {\bf 16} S673

\bibitem{YTO} Hodges J A {\it et al.} 2001 {\it Can. J. Phys.} {\bf 79} 1373

\bibitem{GTO} Bonville P {\it et al.} 2003 {\it J. Phys.: Condens. Matter} {\bf 15} 7777

\bibitem{Champion} Champion J D M {\it et al.} 2003 {\it Phys. Rev. B} {\bf 68} 020401(R)

\bibitem{Poole} Poole A, Wills A S and Lelievre-Berna E 2007 {\it J. Phys.: Condens. Matter} {\bf 19}  452201

\bibitem{Ruff} Ruff J P C {\it et al.} 2008 {\it Phys. Rev. Lett.} {\bf 101} 147205 

\bibitem{Bramwell} S. T. Bramwell \emph{et al.} 2000 {\it J. Phys.: Condens. Matter} {\bf 12}  483

\bibitem{PalmerChalker} Palmer S E and Chalker J T 2000 {\it Phys. Rev. B} {\bf 62} 488

\bibitem{Curnoe} Curnoe S H 2008 {\it Phys. Rev. B} {\bf 78} 094418 

\bibitem{Elhajal} Elhajal M, Canals B and Lacroix C 2004 {\it J. Phys.: Condens. Matter} {\bf 16} S917

\bibitem{Elhajal2} Elhajal M {\it et al.} 2005 {\it Phys. Rev. B} {\bf 71} 094420

\bibitem{Sergienko} Sergienko I A and Curnoe S H 2003 {\it J. Phys. Soc. Jpn.} {\bf 72} 1607

\bibitem{Rosenkranz} Rosenkranz S {\it et al.} 2000 {\it J. Appl. Phys.} {\bf 87} 5914

\bibitem{RareEarths} Jensen J and Mackintosh A R 1991 {\it Rare Earth Magnetism} (Oxford University Press)

\bibitem{Molavian} Molavian H, Gingras M J P and Canals B 2007 {\it Phys. Rev. Lett.} {\bf 98} 157204 


\end{thebibliography}
\end{document}